\documentclass[final]{elsart}

\usepackage{icarus}

\usepackage{natbib}

\usepackage{graphicx}

\usepackage{amssymb}

\usepackage{longtable}

\usepackage{amsmath}

\def\comm#1           {{\tt (COMMENT: #1)}}

\begin{document}

\begin{frontmatter}


\title{AVAST Survey 0.4$-$1.0 $\mu$m Spectroscopy of Igneous Asteroids in the Inner and Middle Main Belt}

\author[Adler]{Michael Solontoi}, 
\author[Adler]{Mark Hammergren},
\author[Adler]{Geza Gyuk},
\author[UAA]{Andrew Puckett}

\address[Adler]{Adler Planetarium, 1300 S. Lake Shore Drive, Chicago, IL 60605, USA}
\address[UAA]{University of Alaska Anchorage, 3211 Providence Drive Anchorage AK 99508}
\footnotesize
Contact Email: msolontoi@adlerplanetarium.org
\normalsize


\end{frontmatter}

\begin{abstract}

We present the spectra of 60 asteroids, including 47 V-types observed during the first phase of the Adler V-Type Asteroid (AVAST) Survey.  SDSS photometry was used to select candidate V-type asteroids for follow up by nature of their very blue $i-z$ color.  47 of the 61 observed candidates were positively classified as V-type asteroids, while an additional six show indications of a 0.9 $\mu$m feature consistent with V-type spectra, but not sufficient for formal classification.  Four asteroids were found to be S-type, all of which had $i-z$ values very near the adopted AVAST selection criteria of $i-z \le -0.2$, including one candidate observed well outside the cut (at a mean $i-z$ of -0.11).  Three A-type asteroids were also identified.  Six V-type asteroids were observed beyond the 3:1 mean motion resonance with Jupiter, including the identification of two new V-type asteroids (63085 and 105041) at this distance.  Six V-type asteroids were observed with low ($<5\deg$) orbital inclination, outside of the normal dynamical range of classic Vestoids, and are suggestive of a non-Vesta origin for at least some of the population.

\end{abstract}


\section {Introduction}


Increasingly evidence shows that the early solar system was a dramatically complex system.  The mechanisms at work in the early Solar System would leave their imprints in the asteroid belt, preserving within the dynamics of this population clues to the late stages of Solar System and giant planet formation\citep{2006Natur.439..821B}.  Tremendous mass loss may have taken place in the vicinity of the asteroid belt, including the loss of protoplanetary cores.  Such large objects would be expected to be differentiated and hence show unique and characteristic surface chemistry, especially when compared to smaller bodies which did not undergo such large scale melting and differentiation.

Spectral studies of Vesta in 1970 indicated a strong signature of the silicate
mineral pyroxene on Vesta's surface, bearing striking similarities to spectra of the basaltic achondrite meteorites \citep{1970Sci...168.1445M}, adding to the evidence of Vesta being perhaps the only remaining remnant of these differentiated planetary cores.  Further work has refined that basic picture, uncovering evidence for surface compositional heterogeneity on a global scale, particularly a distinct mineralogy for surface regions suggestive of non-impact driven evolution, and an enormous crater near its south pole \citep[e.g.][]{ 1997Icar..127..130G, 1997Sci...277.1492T, 2010Icar..208..238L,  2010Icar..210..693R}.  Vesta is  the first \emph{in situ} target of the Dawn spacecraft mission,  which will presumably shed great light on this object's composition and complex geological history \citep{2004P&SS...52..465R}.


The discovery of small ($\sim5-15$ km) apparently basaltic asteroids bridging the gap between Vesta and the 3:1 mean motion resonance with Jupiter has been regarded by many as the demonstrable link between Vesta and the HED meteorites \citep{1993Sci...260..186B}.  Since then more than fifty, small, Vesta-like (taxonomic V-type) asteroids have been spectroscopically confirmed in the inner main asteroid belt \citep[e.g.][]{1995Icar..115....1X, 2002Icar..158..106B, 2004Icar..172..179L}.  Additionally, a dynamic family centered around Vesta includes 4000 known members exhibiting photometric colors that suggest similar surface compositions  \citep{2008Icar..198..138P}.  The majority of these asteroids are likely to be fragments of Vesta, colloquially ``Vestoids," potentially liberated by large crater-forming impacts more than one billion years ago \citep{2008Icar..193...85N}.

The discovery of a basaltic asteroid in the outer main belt, 1459 Magnya, that appears to be 
dynamically unrelated to Vesta \citep{2000Sci...288.2033L} has opened the door to studies of the 
remnants of other differentiated asteroids. In a detailed spectroscopic mineralogical analysis, 
\cite{2004Icar..167..170H} find that Magnya is distinct from Vesta in orthopyroxene chemistry, 
concluding that the compositional difference precludes an origin on Vesta.  

Other studies have found several additional middle and outer main-belt basaltic asteroids, 
including 21238 \citep{2007AAS...210.8907H, 2007LPI....38.1851B, 2007A&A...473..967C},
 10537 \citep{2008ApJ...682L..57M, 2009P&SS...57..229D}, 7472 \citep{2009P&SS...57..229D}, 
 and 40521 \citep{2007A&A...473..967C, 2008Icar..194..125R}.  Because these objects have
 orbital semi-major axes greater than 2.5 AU, and thus reside on the 
other side of the powerful barrier to cross-diffusion that is the 3:1 mean motion resonance with 
Jupiter, many if not most of them are unlikely to be runaways from the Vesta family \citep{2008Icar..193...85N, 2008Icar..194..125R}. These V-type asteroids that appear to be 
unrelated to Vesta  (either dynamically or compositionally) are commonly referred to as Ònon-Vestoids.Ó  

While most of the V-type asteroids in the inner belt are likely to be members of the Vesta dynamical family, it is almost certain that other large protoplanetary cores existed in the region at early times, as evidenced by the diversity of iron meteorites in the meteoritic record \citep{2002aste.conf..697S}.  Even the among the HED meteorites themselves is isotopic inhomogeneity which point to additional parent bodies \citep{2009GeCoA..73.5835S}.  Dynamically this idea is supported in the asteroid belt by the identification of a substantial population of V-type asteroids with low orbital inclinations that are difficult to explain as dynamically evolved collisional members of the Vesta family.  Indeed \cite{2008Icar..193...85N} suggest that a substantial fraction of these low inclination
objects may have an origin independent of Vesta, or may have formed from a different and 
possibly earlier collisional event than the one that formed the main Vesta family.

Several studies have pointed to spectral differences between Vesta family and non-family 
members, although there is some disagreement. \cite{1998AMR....11..163H} find that non-family 
members have steeper visible slopes than family members, which they attributed to an increased 
degree of space weathering, which may also play a role in why the spectra of observed Vestoids tend to be much redder than that of Vesta itself and the HEDs \citep{2011Icar..214..147M}.  While \cite{2011A&A...533A..77D} find an excess of diogenitic material among non-family members \cite{2010Icar..208..773M} find no slope effect in the near-infrared, nor new evidence for mineralogical 
differences between family and non-family members beyond that previously noted for 1459 and 
21238, a disagreement raising the need for additional observations and analysis to be performed on this class of object.  Non-Vestoid 
mineralogies have also been inferred for 7472 and 10537 \citep{2011LPI....42.2483B, 2008Icar..198...77M}. 
The discovery and characterization of additional non-Vestoids is needed to address these 
disagreements, as well as potentially provide clues to the existence and composition of 
differentiated parent bodies independent of Vesta.  The identification of such objects is the major aim of our ongoing observational program, the AVAST (Adler V-type ASTeroid) Survey.


\section{Observation and Reduction}\label{method}

Since 2005 the AVAST survey has conducted a program of visible-to-near-infrared spectroscopic confirmation of asteroids with unusually blue $i-z$ band colors, as measured by the Sloan Digital Sky Survey (SDSS).  In this paper we present the results of the $0.4-1.0\mu$m reflectance spectroscopy of these candidate asteroids using the Dual Imaging Spectrograph (DIS) on the ARC 3.5-m telescope at the Apache Point Observatory.  A future paper will present the results of complementary, ongoing, near infrared reflectance spectroscopy in the $0.9-2.5\mu$m regime.

\subsection{SDSS Target Selection}

The SDSS was a digital photometric and spectroscopic survey that 
covered about one quarter of the Celestial Sphere in the North Galactic cap and a smaller ($\sim
$300 deg$^2$) but much deeper survey in the Southern Galactic hemisphere and began 
standard operations in April 2000 \citep[see][and references within]{2000AJ....120.1579Y, 2002AJ....123..485S, 2009ApJS..182..543A}.  The Seventh SDSS Public Data 
Release \citep{2009ApJS..182..543A} ran through July 2008 and contains over 
357 million unique photometric objects. Of particular interest to Solar 
System studies,  the survey covers the sky at and near the ecliptic from approximately 
ecliptic latitude $\lambda=100^{\circ}$ to $\lambda=225^{\circ}$. The repeat scans of the Southern
Galactic hemisphere (Stripe 82; crossing $\lambda= 0^{\circ}$) also pass through the Ecliptic.

Although designed mainly for observations of extragalactic sources, the SDSS has significantly 
contributed to studies of the solar system, notably in the success it has had with asteroid 
detections, cataloged in the SDSS Moving Object Catalog (hereafter SDSS MOC, \citealt{2002SPIE.4836...98I}). This public, value-added, catalog of SDSS asteroid observations contains, as of its 
fourth release, measurements of 471,000 moving objects, 220,000 of which have been matched 
to 104,000 known asteroids from the ASTORB file\footnote{see ftp://ftp.lowell.edu/pub/elgb/astorb.html.} \citep{2002AJ....124.1776J}. The SDSS MOC data has been 
widely used in recent studies of asteroids  \citep[e.g.][]{2001AJ....122.2749I, 2002AJ....124.1776J,
2007LPI....38.1851B, 2008Icar..198..138P, 2008A&A...488..339A, 2010A&A...510A..43C}.

\begin{figure} [p]
\begin{center}
\includegraphics[width=\textwidth]{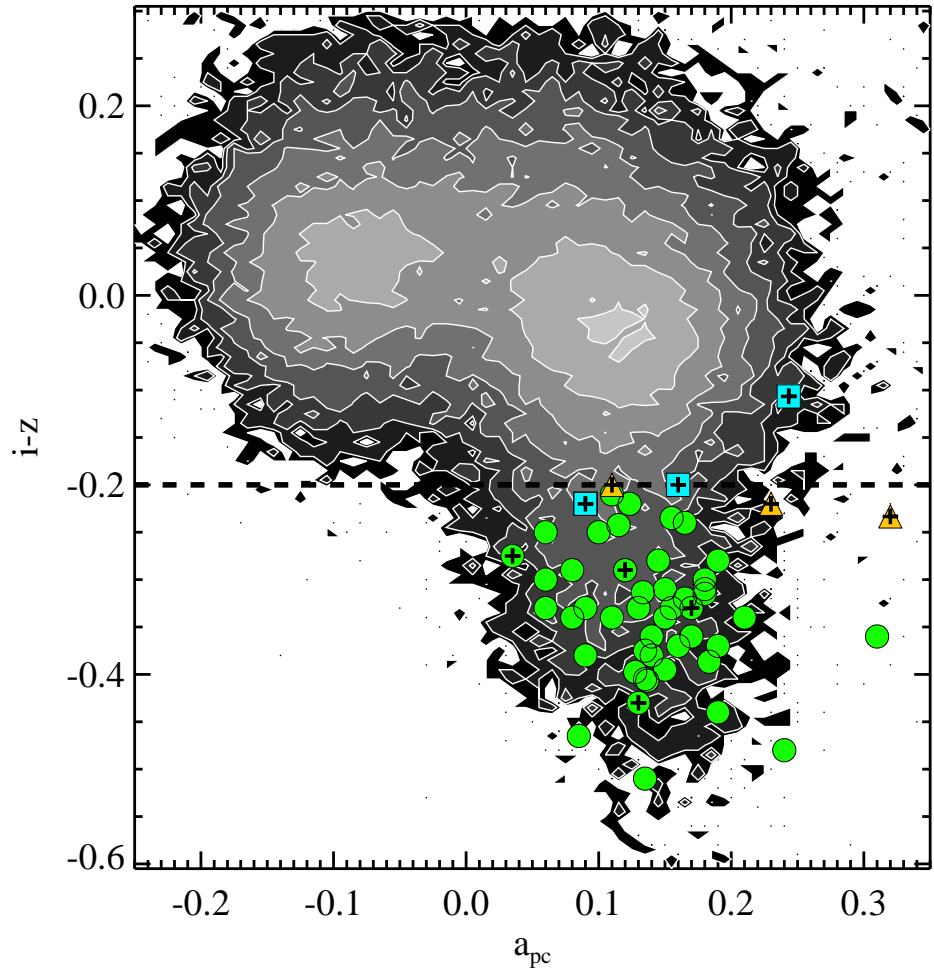} 
\end{center}
\caption[Color-Color Plot]{The a$_{\textrm{pc}}$ vs $i-z$ colors of the asteroids classified in the course of the AVAST survey displaying a significant 0.9 $\mu m$ absorption as seen against the colors of asteroids from SDSS MOC3 (background contours of $2^{n}$).  AVAST V-type asteroids are seen as green circles, A-type asteroids as orange triangles, and S-type as pale blue squares.  Observed asteroids with a semi-major axis greater than 2.5AU are indicated by crosses.  The dashed line represents the main AVAST selection criteria of $i-z < -0.2$.}
\label{fig_color} 
\end{figure}

While the SDSS filters were not specifically chosen for asteroid reflectance studies, they have 
proven to be able to distinguish the major taxonomic types \citep{2001AJ....122.2749I, 2008Icar..198..138P}.  The strong 0.9 $\mu$m absorption features of the V taxonomic types lies within the $z$-band (centered at 8931\AA).  This absorption feature produces very blue $i-z$ band colors relative to asteroids without such an absorption feature (e.g. C-types) or a weak one (e.g. S-types).  A color-color plot of the MOC3 asteroids in $i-z$ and the principle component color $a_{pc}$, with $a_{pc}$ defined by \citealt{2001AJ....122.2749I} as $a_{pc} = 0.89(g-r) + 0.45(r-i) -0.57$, clearly shows a highly bimodal distribution of C-types at $a_{pc}\sim-0.1$, and S-types at $a_{pc}\sim0.15$ (See Figure \ref{fig_color}).  Those asteroids consistent with a V-type taxonomy form a population below the S-types at bluer $i-z$ colors and allow for selection via a simple color cut.

For AVAST, we have selected asteroids with $i-z \le -0.2$ as shown in Figure \ref{fig_color}. It is important to note that our intent 
never has been to produce an unbiased sampling across the main belt. We have preferentially 
selected non-Vesta family targets, focusing our efforts on those in interesting dynamical locations 
(e.g. in the middle and outer belt, or at inclinations significantly less than that of 
Vesta).  As such this target selection is independent of similar SDSS selection criteria used by other studies (e.g. 
\cite{2006Icar..183..411R}, \cite{2007LPI....38.1851B}, and \cite{2008Icar..198...77M}) though 
naturally there exists overlap in target lists.  

This criteria was applied to the Third Release of the SDSS MOC  which contains data on 204,305 
moving objects, including astrometric and photometric observations of 43,424 previously 
known asteroids \citep{2002AJ....124.2943I}.  While the Third Release SDSS MOC data set has 
been suplanted by the Fourth Release, it remains the ``gold standard'' for photometric quality, as 
the fourth release includes SDSS-II data \citep{2009ApJS..182..543A} obtained in non-photometric conditions, requiring additional processing to assure robust, uniform data 
quality (see \citealt{2008Icar..198..138P} for details).  In order to keep our selection criteria 
uniform through this work we choose to focus on our initial set of candidates chosen from the 
third release of the SDSS MOC, with additional candidates selected with quality control requirements from the fourth release. The color-color plot of asteroids observed by AVAST are seen against the membership of the SDSS MOC3, along with the AVAST color-cut in $i-z$ in Figure \ref{fig_color}.

We have observed the spectra of 61 asteroids selected by our criteria.  The majority ($\sim77$\% of our targets have proven to have classifiable spectra with V-type asteroids based on the feature-based taxonomy of \cite{2002Icar..158..146B}.  Table \ref{table_obs} presents a list of the observed asteroids.  Furthermore all but one observed asteroid show the 0.9 $\mu$m absorption feature characteristic of a thermally metamorphosed surface mineralogy from highly processed V-types to less processed members of the S-type complex.  The single case where our selection criteria failed to select an asteroid with a prominent 0.9 $\mu$m feature, asteroid 44496 (1998 XM5) can be traced to the asteroid's photometric position being coincident with an artificial satellite trail in the SDSS $i$-band image.  This false positive prompted the inclusion of SDSS photometric processing flags \citep{2002AJ....123..485S} into our revised selection criteria to exclude such cases.  This C-type asteroid 44496 is not included in the discussion of the AVAST results.

\subsection{Spectroscopy of Candidate V-type asteroids}

The observations were performed from 2005-2007 at the Apache Point
Observatory, using the Dual Imaging Spectrograph (DIS) on the Astrophysical Research 
Corporation 3.5m telescope.

The DIS uses two cameras to simultaneously record the blue and red spectral regions. The low resolution blue grating and medium resolution red grating provided dispersions of 2.42 and 2.31 \AA\ pixel$^{-1}$, respectively.  During the summer of 2006 the instrument's gratings were replaced resulting in a blue dispersion of 1.83 \AA\ pixel$^{-1}$.  This configuration permits the coverage of the spectral range from approximately $0.36-1.0$ $\mu$m.  The dichroic mirror has a transition at approximately 0.55 $\mu$m, causing strong variations in throughput extending to about 0.25 $\mu$m on either side. These variations are imperfectly removed in reduced spectra, so the spectral region immediately around 0.55 $\mu$m is excluded from the plots of asteroid spectra. 

The 1.5-arcsecond wide spectrograph slit was maintained at the parallactic angle to minimize the effects of differential refraction. Solar analog stars were observed periodically to remove telluric absorptions and for production of the reflectance spectrum.  These solar analogs were chosen to match the airmass at which each asteroid was observed.

Data reduction was performed using the Image Reduction and Analysis Facilty (IRAF) package, 
following standard procedures. After subtraction by an average bias frame and division by an 
average flat field, the spectra were extracted. Wavelength calibration was performed using 
observations of helium, neon, and argon arcs. The spectrum of 
each asteroid was divided by the spectrum of the solar analog star at similar airmass. Finally, the 
reflectance spectrum was normalized to unity at 0.55 $\mu$m by convention, using a linear fit between the red and blue sides across the dichroic region.

\section {Results}\label{result}

\subsection{Visible Spectra of Candidates}\label{sec_spec}

\begin{figure} [p]
\begin{center}
\includegraphics[width=\textwidth]{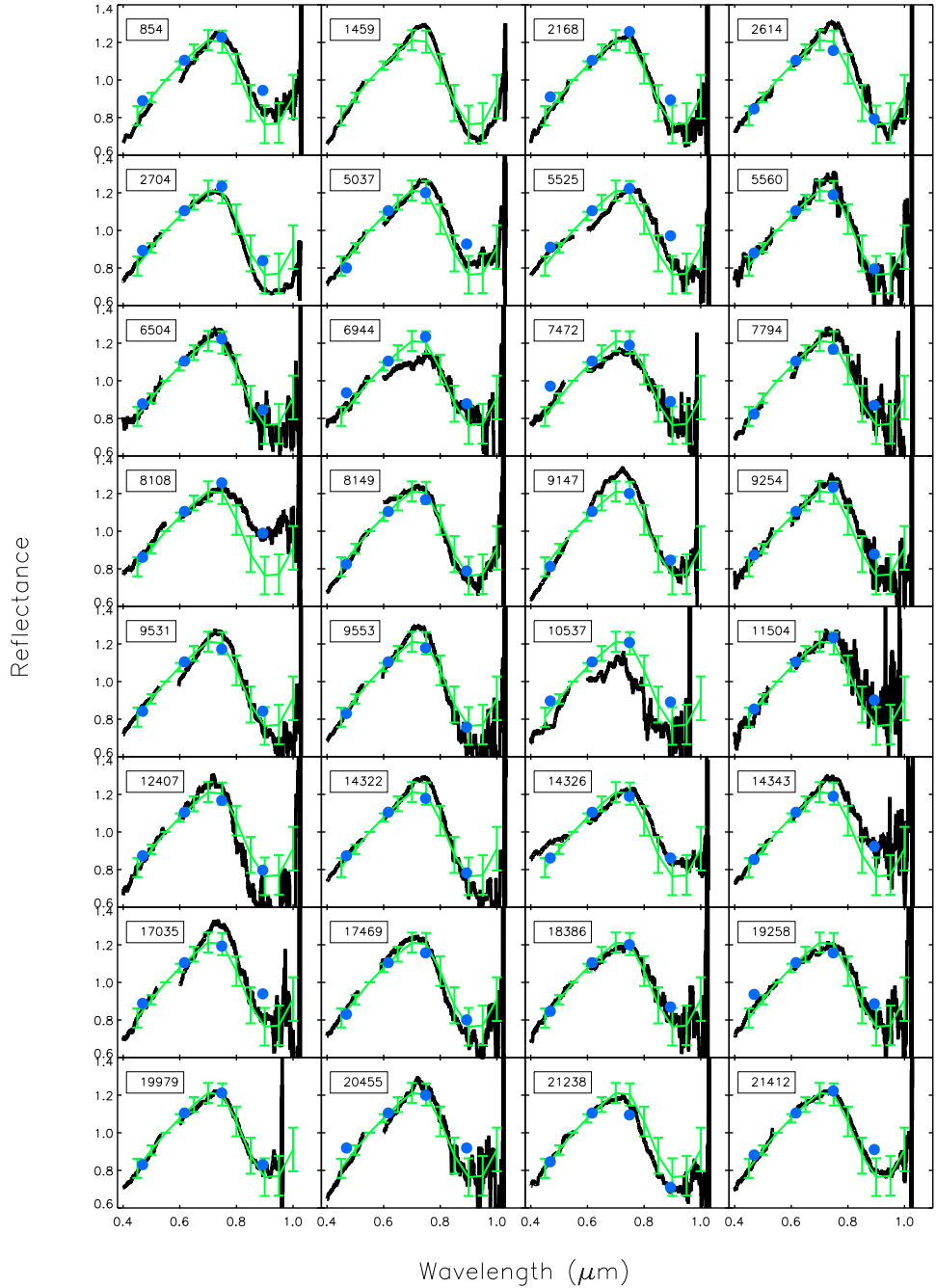} 
\end{center}
\caption[grids of spectra]{Reflectance spectra of the asteroids observed by AVAST with significant 0.9 $\mu$m absorption feature consistent with V,A, or S-type taxonomy.  The data (black lines) has been median filtered, and normalized to 0.55 $\mu$m by convention, with the \cite{2002Icar..158..146B} V-type taxonomy template overplotted in green.  The blue dots show the normalized reflectance from the SDSS {\it g,r,i,z} bands.}
\label{fig_grid} 
\label{lastfig}
\end{figure}

Because our spectral coverage terminates near 1 $\mu$m, our taxonomic classifications are 
based on the visible wavelength SMASS II / Bus taxonomy \citep{2002Icar..158..146B}. Our 
spectra continue to provide useful information in the near-infrared to around 1 $\mu$m, giving us 
more coverage of the 0.9 $\mu$m feature than was available to SMASS II. This permits the use of the band centers and band widths as aids in our classification.  Using DIS, to date, we have observed 47 objects which have spectra consistent 
with the taxonomic class V and three asteroids with A-type spectra, potentially pointing to an igneous nature.  A-type asteroids, while fragments of differentiated asteroids, lack the deep pyroxine absorption at 0.9 $\mu$m, but have fairly blue $i-z$ colors due to a nearly uniformly declining slope in the $z$-band due to their olivine chemistry.  An 
additional four asteroids are best matched to S-complex types in the \cite{2002Icar..158..146B}, 
showing a less deep 0.9 $\mu$m absorption feature, and another six asteroids, which show a deep 0.9 $\mu$m feature, but can not be robustly classified with the \cite{2002Icar..158..146B} taxonomy schema, producing a suggested classification consistent with several spectral types.  Table \ref{table_obs} presents a list of the observing geometry, orbital elements and \cite{2002Icar..158..146B} taxonomy of all of the 
60 observed asteroids with a 0.9 $\mu$m absorption feature, including a best estimate from visual inspection of the 6 that failed the taxonomic schema.  Each asteroid's spectra is shown in Figure \ref{fig_grid}, along with the normalized reflectance derived from the asteroid's {\it g,r,i,z} photometry using the solar transformations given in \cite{2001AJ....122.2749I}.  1459 Magnya does not have an observation in the third release of the MOC, but was selected for observation as it was a known V-type beyond 2.5AU.

\subsection{Igneous asteroids dynamically isolated from Vesta}

\begin{figure} [p]
\begin{center}
\includegraphics[width=\textwidth]{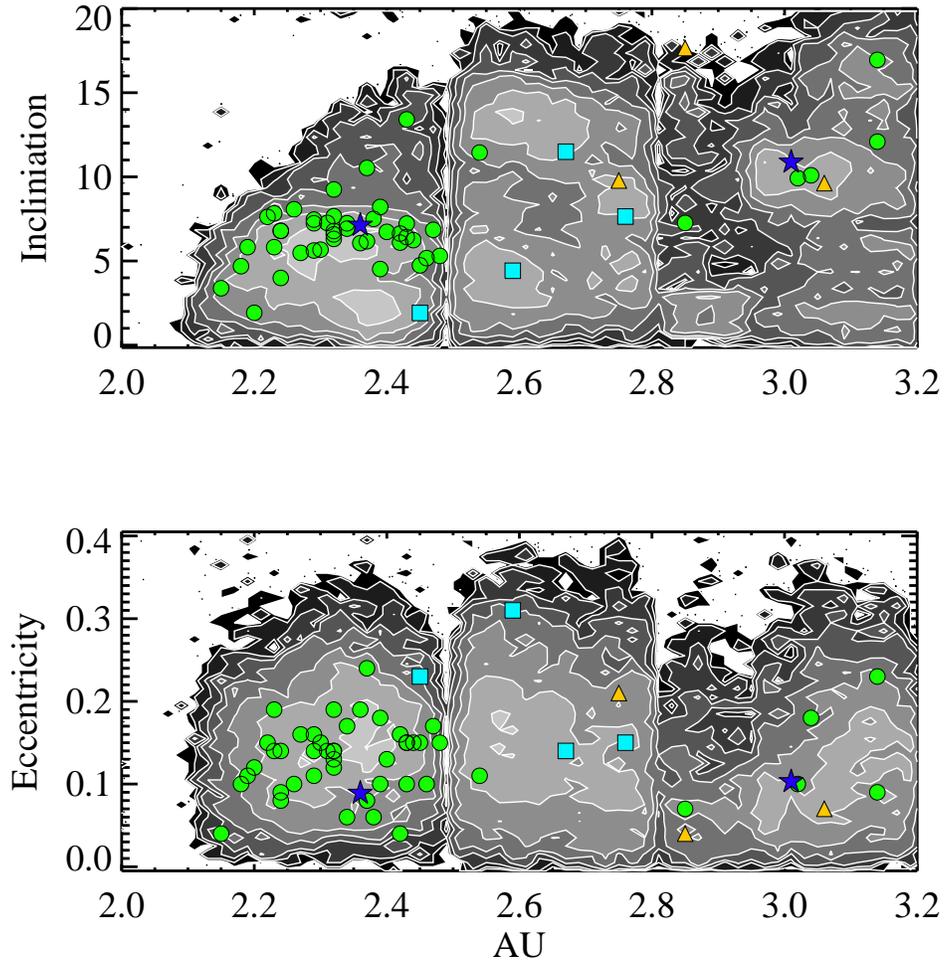} 
\end{center}
\caption[Orbit diagram for the AVAST asteroids]{The semi major axis vs. inclination (top) and eccentricity (bottom) of the asteroids classified in the course of the AVAST survey displaying a significant 0.9 $\mu$m absorption feature as seen agains elements of known asteroids from SDSS MOC3 (background contours of $2^{n}$). AVAST V-type asteroids shown as green circles, A-type as orange triangles, and S-type as light blue triangles.  The location of Vesta and Eos are indicated by dark blue stars.}
\label{fig_orbit} 
\end{figure}

AVAST has classified six V- and three A-type asteroids with orbital semi-major axes greater 
than 2.5 AU, and eight V-types with inclinations outside $3\sigma$ of the \cite{2008Icar..198..138P} 
defined Vesta family, Six of which are below the lower ($<5^{\circ}$) inclination limit.  In particular these 
objects at large semi-major axis, or at low inclination have a significant probability of being 
independent of Vesta \citep{2008Icar..193...85N,2008Icar..194..125R} which affords us the 
chance to study now-missing parent bodies of differentiated asteroids. It is especially interesting 
to note one apparently igneous asteroid in the vicinity of Eos and its associated family of 
asteroids, which may represent fragments of a partially differentiated chondritic body \citep
{2008Icar..195..277M}.  Figure \ref{fig_orbit} shows the orbital distribution of the asteroids with 
deep V- or A-type 0.9 $\mu$m absorption feature.  

Of particular note are those objects with a $>$ 2.5 AU, i.e., on the other side of the 3:1 mean 
motion resonance from the Vesta family.  We independently selected and confirmed a basaltic 
nature for the middle and outer main belt asteroids 1459, 7472, 10537, and 21238.  In addition 
to these objects, we find that asteroids 63085 and 105041 also display strong 0.9 $\mu$m 
absorption features indicative of basalt and with visible spectra consistent with a V-type 
classification.

While mineralogical implications of V-type spectra are more rigorous with both the 0.9 $\mu$m and 2.0 $\mu$m bands for spectral discrimination \citep{2009Icar..202..160D}, the 0.9 $\mu$m band centers may be studied for evidence of statistical differences between asteroids.  These band centers were found by fitting the linear slope and the 0.9 $\mu$m band through the use of a Modified Gaussian Model \citep{1990JGR....95.6955S} to fit each spectra with a combination of linear and Gaussian fits.  We group the classified V-type asteroids in the study into three populations, a ``classic" dynamical Vestoid population, those at low ($<5^{\circ}$) inclination and those beyond the 3:1 Jovian resonance.  We find that both the low inclination and the distant populations of V-types have band centers shifted to slightly shorter wavelengths, with the classic Vestoids having a mean band center ($\pm 1\sigma$) at $0.921\pm0.009\mu$m, the low inclination group at $0.918\pm0.011\mu$m and the distant population at $0.907\pm0.019\mu$m.  To characterize these differences a Kolmogorov-Smirnov test (KS-test) was performed on the band center distributions of the populations in question. Between the classic Vestoids and the low inclination population, the KS-test probability value was 0.77, suggesting the low inclination population have been drawn from the same population as the classical Vestoids.  The KS-test probability value for the band center distributions between the classical Vestoids, and those beyond 2.5AU however, is only 0.10, suggestive of being unlikely that these two groups share a common population as their source. 

These results stand in contrast with the findings of \cite{2010Icar..208..773M}, who find their low inclination V-type populations 0.9 $\mu$m band centers trend toward longer wavelengths.  The band centers of the low inclination are virtually the same as the main classical Vestoid population here.  In particular, adapting the low inclination definition of $<6^{\circ}$ used by \cite{2010Icar..208..773M}, the band centers shift even closer together, with a mean of 0.918 $\pm$ 0.010$\mu$m for the Vestoids, and of  0.919  $\pm$  0.011$\mu$m for the low inclination population, and a KS-test probability value of 0.98.

\subsection{A-type asteroids}

Differentiation is generally believed to result in an object with a basaltic crust, olivine-dominated mantle, and nickel-iron core. Although more than one hundred basaltic asteroids and dozens of apparently metallic asteroids have been identified in the main belt, only a relative handful of asteroids have been found to bear the distinctive signature of a dominant olivine composition (corresponding to taxonomic type A). This lack of mantle material among the asteroids is mirrored by the lack of olivine-dominated asteroidal meteorites. This unusual paucity has long been considered a serious problem in our understanding of the asteroid belt (see \cite{1996M&PS...31..607B} for a review of the subject). 
A prominent hypothesis is that the disruption of differentiated planetesimals occurred so early in the history of the solar system, and that the subsequent collisional evolution of the asteroids was so extensive, that fragments of the mantle and crust material of those objects have been ground down to sizes below observable limits, or Òbattered to bitsÓ \citep{1996M&PS...31..607B}. If this is the case, then there may be a population of small, olivine-rich asteroids that simply have not yet been identified and characterized. 
We have classified three A-type asteroids in AVAST, all of which have orbits beyond 2.5 AU. Figure \ref{fig_orbit} shows the orbital distribution of these A-type asteroids. These discoveries, enabled by the vast increase in the numbers of known asteroids since 1996, combined with our ability to selectively target the relatively rare A- and V-type asteroids through their SDSS colors, may imply that our survey is reaching down to the size of the larger fragments of the population described by the Òbattered to bitsÓ model. Alternatively, these small A-type asteroids, along with the V-type asteroids that appear to be dynamically unrelated to Vesta, may represent fragments of differentiated bodies scattered outward from the terrestrial planet region \citep{2006Natur.439..821B}.

%

\section {Discussion}\label{sum}

The results of the AVAST survey add additional weight to the use of SDSS photometric colors in selecting populations of asteroids for follow up work, in particular the use of $i-z$ color to select candidate V-type asteroids.  47 of the 61 observed candidates were positively classified as V-type asteroids, while an additional six show indications of a 0.9 $\mu$m feature consistent with V-type spectra, but of insufficient quality for formal classification.  Four asteroids were formally classified as part of the S-type complex, all of which had $i-z$ values very near the adopted AVAST selection criteria of $i-z \le -0.2$, including one candidate observed well outside the cut.  This asteroid, 46262,  was observed three times by the epoch of MOC3, and showed a wide range of measured SDSS $i-z$ colors, from -0.48 to 0.13.  The relatively low photometric errors ($\sigma\_{i-z} < 0.1$) combined with its semi major axis of 2.76 AU, and potential for longitudinal color variation warranted its inclusion in the survey.  AVAST spectroscopy of this object was consistent with an S-type.  Adopting a more conservative color cut of $i-z \le -0.25$ removes all of the confirmed S-types from the survey, leaving a sample of only V-types among the spectroscopically classified asteroids by AVAST.  While doing so would eliminate false positives from the V-type sample, it would also eliminate the serendipitously selected A-type asteroids, which are also thought to come from differentiated asteroids.  Indeed in the SDSS $z$-band the almost linearly declining olivine absorption feature of A-type asteroids and the less pronounced 0.9 $\mu$m feature of the S-types produce a similar $i-z$ colors, and any conservative color cut made to remove outlier S-types would strip the sample of A-type asteroids as well.  Regardless of which cut is used, this high rate of successful identification of V-type asteroids from SDSS colors (77\% with our original cut, 100\% with a more conservative one) bodes well for the future study of asteroids selected from current and future large scale surveys based on well calibrated photometric colors.

The meteoritic record, particularly the diversity of iron meteorites, provides evidence for the former presence of large, differentiated, parent bodies.  AVAST has revealed a significant amount of dynamically separated igneous material through the identification of V- and A-types asteroids in non-Vestoid orbits.  One such population found are low inclination V-types which are difficult to link dynamically to the current family of Vestoids, potentially pointing to a non-Vesta origin, or a past orbit of Vesta significantly different than today \citep{2008Icar..193...85N}.  Even more suggestive of being fragments of non-Vesta parent bodies are the V-types beyond the 3:1 mean motion resonance with Jupiter at 2.5 AU.  Studies of the dynamical evolution of the Vestoids compared to individual V-types beyond 2.5 AU show a high improbability that these distant asteroids could be ``run-away" members of the Vesta family \citep{2008Icar..193...85N, 2008Icar..194..125R}.  The largest of these distant V-types, 1459 Magnya, has been shown to have surface chemistry distinct from that of the Vestoids \citep{2004Icar..167..170H}.  The additional V-types identified by this, and other spectroscopic surveys, are smaller and fainter, and will require further high quality near-infrared spectra to determine whether their pyroxine chemistries are distinct from Vesta, and would serve to strengthen the dynamical case for their independent origin.

The growing number of confirmed V-types beyond 2.5 AU lends weight to an argument for a non-Vesta origin for many of them.  They are still lacking, however, in the number of bodies needed to potentially identify family membership, of either known or, as of yet unknown, dynamical families.  Although three of the objects studied in our survey (7472, 27202, and 105041) appear to be clumped near the Eos family around 3.0 AU, only one of them (27202) was found to be a member of the Eos family by \cite{2010PDSS..133.....N}. An examination of the proper elements of asteroids in that region (Figure \ref{fig_orbit}) shows that asteroids 27202 and 105041 are on the outskirts of the Eos family, both having proper eccentricities higher than the bulk of the Eos family. However, this region is threaded with numerous mean motion, secular, and three-body resonances \citep{2006Icar..182...92V}, which may have helped to potentially drive these asteroids away from the family.

While a study of the 0.9 $\mu$m band center shows a statistical dissimilarity between the observed distant V-types (beyond 2.5 AU), and those in the classical dynamic region of the Vestoids, we find no such difference between that same Vestoid population, and those asteroids with low orbital inclinations.   A more rigorous classification of asteroid spectra is available to those with both visual and near-infrared spectroscopy.  Particularly for V-type asteroids the addition of the near-infrared spectra adds the $\sim$ 2.0 $\mu$m absorption, and combined with the 0.9 $\mu$m feature allows for a much more robust classification \citep{2009Icar..202..160D}, and variant mineralogies may be determined by the analysis of both these spectral regions.  In some cases the reliance on only the visible spectra from 0.4 - 1.0 $\mu$m may also lead to mis-identification.  The reflectance spectrum of 7472 Kumakiri shows a classic V-type 0.9 $\mu$m absorption feature, leading this work to classify it based on the visible spectrum as a V-type.  \cite{2011LPI....42.2483B} has shown that 7472 does not match a V-type spectra template in the near infrared, lacking the corresponding V-type absorption feature at $\sim$ 2.0 $\mu$m, but rather resembles the O-type asteroid 3628 Boznemcova.  With this in mind, the next phase of the AVAST survey, currently underway, is a complimentary NIR survey of these asteroids with the goal of finding evidence of independent mineralogy in addition to dynamical independence for the asteroids outside of the Vesta family.

\section*{Acknowledgments}

The authors gratefully acknowledge support from NASA Planetary Astronomy Grant \#NNG06GI40G and the Brinson Foundation grant in aid of Astrophysics Research to the Adler Planetarium \& Astronomy Museum, and would like to thank Nicholas Moskovitz and an anonymous referee for their insightful comments and suggestions.

Work here based on observations obtained with the Apache Point Observatory 3.5-meter telescope, which is owned and operated by the Astrophysical Research Consortium.



\label{lastpage}

\newcommand{\aj}{Astrophys. J.}
\newcommand{\apjs}{ApJS}
\newcommand{\iaucirc}{IAU Circ.}
\newcommand{\aap}{A\&A}
\newcommand{\apj}{ApJ}
\newcommand{\mnras}{MNRAS}
\newcommand{\jgr}{JGR}
\newcommand{\nat}{Nature}
\newcommand{\icarus}{Icarus}
\newcommand{\planss}{P\&SS}
\newcommand{\ajfl}{ApJL}

\bibliographystyle{apj}

\small
\bibliography{AVAST_bib.bib}
\normalsize


\clearpage



{
\renewcommand{\baselinestretch}{1}
\small\normalsize

\begin{center}
\begin{longtable}{|l|c|c|c|c|c|c|c|c|c|c|}

\hline   \hline Asteroid 	&	Date	 &	V	&	$\phi$	&	H	&	a	&	e	&	i	&	Tax  & Band & Solar \\ \hline 
\endfirsthead

\hline   Asteroid	&	Date	&	V	&	$\phi$	&	H	&	a	&	e	&	i	&	Tax  & Band & Solar \\ \hline 
\endhead



854	&	20060101	&	16.3	&	9.6	&	11.9	&	2.37	&	0.17	&	6.09	&	*	&	0.919	&	BD+00\_2717	$^{2}$	\\
1459	&	20060101	&	15.1	&	10.9	&	10.3	&	3.14	&	0.23	&	16.95	&	V	&	0.925	&	HD120528	$^{1}$	\\
2168	&	20060204	&	17.7	&	23.6	&	12.5	&	2.45	&	0.15	&	4.74	&	V	&	0.912	&	HD28099	$^{1}$	\\
	&	20061216	&	17.6	&	17.5	&	12.5	&	2.45	&	0.15	&	4.74	&	V	&	0.919	&	HD120528	$^{1}$	\\
2614	&	20070108	&	16.9	&	12.1	&	13.3	&	2.34	&	0.17	&	6.92	&	V	&	0.927	&	HD120528	$^{1}$	\\
2704	&	20061216	&	16.2	&	14.5	&	13.2	&	2.39	&	0.1	&	4.52	&	V	&	0.921	&	HD120528	$^{1}$	\\
5037	&	20060101	&	15.9	&	12.1	&	13.1	&	2.27	&	0.11	&	7.03	&	*	&	-	&	BD+00\_2717	$^{2}$	\\
	&	20060126	&	16.6	&	22.2	&	13.1	&	2.27	&	0.11	&	7.03	&	*	&	0.913	&	HD28099	$^{1}$	\\
5525	&	20060101	&	16.8	&	12.6	&	13	&	2.22	&	0.15	&	7.62	&	V	&	0.940	&	HD120528	$^{1}$	\\
5560	&	20070108	&	17.5	&	26.1	&	13.6	&	2.29	&	0.11	&	5.62	&	V	&	-	&	HD120528	$^{1}$	\\
6504	&	20070108	&	17.6	&	17.5	&	13.8	&	2.42	&	0.16	&	6.09	&	V	&	0.923	&	HD120528	$^{1}$	\\
6944	&	20060204	&	17.8	&	5.5	&	14.1	&	2.32	&	0.14	&	7.66	&	V	&	0.938	&	HD28099	$^{1}$	\\
7472	&	20061025	&	17	&	17.6	&	12	&	3.02	&	0.1	&	9.92	&	V	&	0.934	&	HD28099	$^{1}$	\\
	&	20061216	&	15.8	&	2	&	12	&	3.02	&	0.1	&	9.92	&	V	&	-	&	P177D	$^{3}$	\\
7794	&	20060101	&	17.2	&	8.4	&	13.6	&	2.3	&	0.15	&	5.67	&	V	&	0.912	&	BD+00\_2717	$^{2}$	\\
8108	&	20060204	&	17.4	&	1.2	&	14	&	2.37	&	0.12	&	6.74	&	*	&	0.912	&	HD28099	$^{1}$	\\
8149	&	20060425	&	16.5	&	15.9	&	13.2	&	2.32	&	0.14	&	6.58	&	V	&	0.922	&	HD191854	$^{1}$	\\
9147	&	20051224	&	16.5	&	13.7	&	13.7	&	2.19	&	0.11	&	5.82	&	V	&	0.915	&	HD120528	$^{1}$	\\
9254	&	20070108	&	17.5	&	17.3	&	13.8	&	2.36	&	0.19	&	6.06	&	V	&	0.939	&	HD120528	$^{1}$	\\
9531	&	20060101	&	17.2	&	9.4	&	13.7	&	2.23	&	0.19	&	5.82	&	V	&	0.921	&	HD120528	$^{1}$	\\
9553	&	20060101	&	17.2	&	11.6	&	14.6	&	2.2	&	0.12	&	1.92	&	V	&	0.912	&	HD120528	$^{1}$	\\
10537	&	20051104	&	18.3	&	19.7	&	12.4	&	2.85	&	0.07	&	7.26	&	V	&	0.880	&	HD144873	$^{1}$	\\
	&	20061025	&	17.4	&	6.3	&	12.4	&	2.85	&	0.07	&	7.26	&	V	&	0.916	&	HD28099	$^{1}$	\\
11504	&	20061025	&	16	&	11.4	&	13.9	&	2.29	&	0.16	&	7.24	&	V	&	0.912	&	BD+00\_2717	$^{2}$	\\
12407	&	20060603	&	18.3	&	20.3	&	14.3	&	2.4	&	0.13	&	6.73	&	V	&	0.901	&	P330E	$^{3}$	\\
14322	&	20060101	&	16.8	&	12	&	13.8	&	2.23	&	0.14	&	7.83	&	V	&	0.920	&	HD120528	$^{1}$	\\
14326	&	20061216	&	16.2	&	3.3	&	14.2	&	2.47	&	0.17	&	6.85	&	V	&	0.915	&	HD120528	$^{1}$	\\
14343	&	20060101	&	16.7	&	3.7	&	13.7	&	2.24	&	0.14	&	6.79	&	V	&	0.916	&	BD+00\_2717	$^{2}$	\\
17035	&	20060127	&	17.2	&	5.5	&	13.6	&	2.44	&	0.15	&	6.24	&	V	&	0.913	&	HD28099	$^{1}$	\\
17469	&	20060425	&	16.1	&	8.5	&	13.2	&	2.37	&	0.08	&	6.17	&	V	&	0.920	&	HD191854	$^{1}$	\\
18386	&	20060425	&	17.4	&	3.9	&	15.1	&	2.27	&	0.16	&	5.48	&	V	&	0.934	&	HD191854	$^{1}$	\\
19258	&	20060603	&	17	&	9.3	&	14.6	&	2.29	&	0.14	&	7.46	&	V	&	0.930	&	P330E	$^{3}$	\\
19979	&	20051224	&	16.4	&	20.6	&	12.6	&	2.46	&	0.1	&	5.17	&	V	&	0.902	&	HD120528	$^{1}$	\\
20455	&	20060425	&	17.5	&	15.4	&	14.6	&	2.32	&	0.13	&	6.78	&	V	&	0.908	&	HD191854	$^{1}$	\\
21238	&	20050415	&	16.4	&	5.4	&	13.1	&	2.54	&	0.11	&	11.44	&	V	&	0.908	&	HD144873	$^{1}$	\\
21412	&	20070108	&	17.1	&	6.2	&	14.7	&	2.15	&	0.04	&	3.38	&	V	&	0.931	&	HD120528	$^{1}$	\\
22759	&	20060204	&	17.5	&	27.2	&	13.4	&	2.39	&	0.18	&	8.21	&	V	&	0.902	&	HD28099	$^{1}$	\\
25327	&	20060122	&	17.3	&	11.2	&	14.2	&	2.43	&	0.15	&	13.41	&	V	&	0.908	&	BD+00\_2717	$^{2}$	\\
27202	&	20060524	&	17.5	&	16.6	&	12.9	&	3.06	&	0.07	&	9.62	&	A	&	-	&	HD120528	$^{1}$	\\
27437	&	20051224	&	16.9	&	3.9	&	13.9	&	2.31	&	0.14	&	7.26	&	V	&	0.927	&	BD+00\_2717	$^{2}$	\\
28256	&	20051224	&	16.6	&	5	&	13.8	&	2.34	&	0.06	&	7.24	&	V	&	0.907	&	HD120528	$^{1}$	\\
28291	&	20051224	&	16.6	&	17.9	&	13.1	&	2.43	&	0.1	&	7.22	&	V	&	0.918	&	HD28099	$^{1}$	\\
29550	&	20060127	&	17.7	&	25.5	&	13.3	&	2.69	&	0.2	&	13.87	&	*	&	-	&	HD28099	$^{1}$	\\
30282	&	20060127	&	17.8	&	17.1	&	14.5	&	2.32	&	0.12	&	6.31	&	V	&	0.905	&	HD28099	$^{1}$	\\
30802	&	20060127	&	17.3	&	0.7	&	14.1	&	2.45	&	0.23	&	1.91	&	S	&	-	&	HD28099	$^{1}$	\\
31584	&	20060127	&	17.2	&	9.6	&	14.7	&	2.38	&	0.06	&	7.5	&	V	&	0.917	&	HD28099	$^{1}$	\\
31692	&	20060127	&	17.3	&	10.5	&	14	&	2.42	&	0.04	&	6.63	&	V	&	0.906	&	HD28099	$^{1}$	\\
32272	&	20060122	&	17	&	1.3	&	14.6	&	2.4	&	0.06	&	5.59	&	*	&	0.929	&	HD144873	$^{1}$	\\
33490	&	20060204	&	17.9	&	3.5	&	14.2	&	2.43	&	0.15	&	6.42	&	V	&	0.926	&	HD28099	$^{1}$	\\
35062	&	20060122	&	16.7	&	7	&	14.2	&	2.37	&	0.24	&	10.53	&	V	&	0.922	&	BD+00\_2717	$^{2}$	\\
36767	&	20060603	&	18.5	&	25.3	&	15.4	&	2.18	&	0.11	&	5.82	&	*	&	0.922	&	P330E	$^{3}$	\\
37705	&	20060603	&	18.2	&	21.6	&	15.3	&	2.18	&	0.1	&	4.69	&	V	&	0.917	&	P330E	$^{3}$	\\
43964	&	20061216	&	16.5	&	5.7	&	14.3	&	2.24	&	0.09	&	6.79	&	V	&	0.923	&	BD+00\_2717	$^{2}$	\\
45708	&	20060718	&	18.1	&	4.5	&	14.2	&	2.76	&	0.15	&	7.64	&	S	&	-	&	SAO103395	$^{1}$	\\
46262	&	20050415	&	18.4	&	21	&	14	&	2.67	&	0.14	&	11.49	&	S	&	-	&	HD144873	$^{1}$	\\
46690	&	20070108	&	17.3	&	4.3	&	15.1	&	2.24	&	0.08	&	3.99	&	V	&	0.936	&	HD120528	$^{1}$	\\
48629	&	20061216	&	17.3	&	17.4	&	14.6	&	2.32	&	0.19	&	9.25	&	V	&	0.914	&	HD120528	$^{1}$	\\
52726	&	20051106	&	17.5	&	2.8	&	13.4	&	2.85	&	0.04	&	17.65	&	A	&	-	&	HD28099	$^{1}$	\\
63085	&	20051106	&	18.2	&	3.7	&	14.2	&	3.14	&	0.09	&	12.09	&	V	&	0.888	&	HD28099	$^{1}$	\\
67652	&	20060425	&	18.1	&	11.7	&	15.7	&	2.26	&	0.1	&	8.07	&	V	&	0.910	&	HD191854	$^{1}$	\\
81542	&	20051106	&	16.6	&	5.2	&	14.5	&	2.59	&	0.31	&	4.42	&	S	&	-	&	HD28099	$^{1}$	\\
93250	&	20060204	&	17.9	&	3.4	&	14.8	&	2.48	&	0.15	&	5.3	&	V	&	0.898	&	HD28099	$^{1}$	\\
105041	&	20060718	&	18.4	&	15.6	&	14.2	&	3.04	&	0.18	&	10.09	&	V	&	0.899	&	HD11532	$^{4}$	\\
129474	&	20060524	&	19.1	&	9.3	&	15.4	&	2.75	&	0.21	&	9.78	&	A	&	-	&	HD120528	$^{1}$	\\
\hline \hline
\caption[Observed Asteroids]{Table of AVAST asteroid observations, with the asteroid number, date of the observation (yyyymmdd), the estimated V magnitude and phase angle ($\boldsymbol{\phi}$) at the time of observation, the absolute magnitude (\textbf{H}), the orbital semi-major axis (\textbf{a}), eccentricity (\textbf{e}), inclination (\textbf{i}), assigned \cite{2002Icar..158..146B} taxonomy (\textbf{Tax}), measured 0.9 $\mu$m band center (\textbf{Band}) in $\mu$m, and comparison solar analog star.  (\textbf{*}) symbols indicate that the spectra display a strong 0.9$\mu$m silicate feature, but fails a rigorous classification, and correspond to the unclassified asteroids in Section \ref{sec_spec}.  S-type here refers to spectra consistent with the S-complex of taxonomic types rather than the specific S-type.  Solars were chosen from $^{1}$\cite{1978A&A....63..383H}, $^{2}$\cite{1992AJ....104..340L}, $^{3}$\cite{1997AJ....113.1138C}, and $^{4}$\cite{2007A&A...471..331D}. }
\label{table_obs}
\label{lasttable}
\end{longtable}
\end{center}
\normalsize

\normalsize
}

\clearpage


\begin{figure} [p]
\begin{center}
\includegraphics[width=\textwidth]{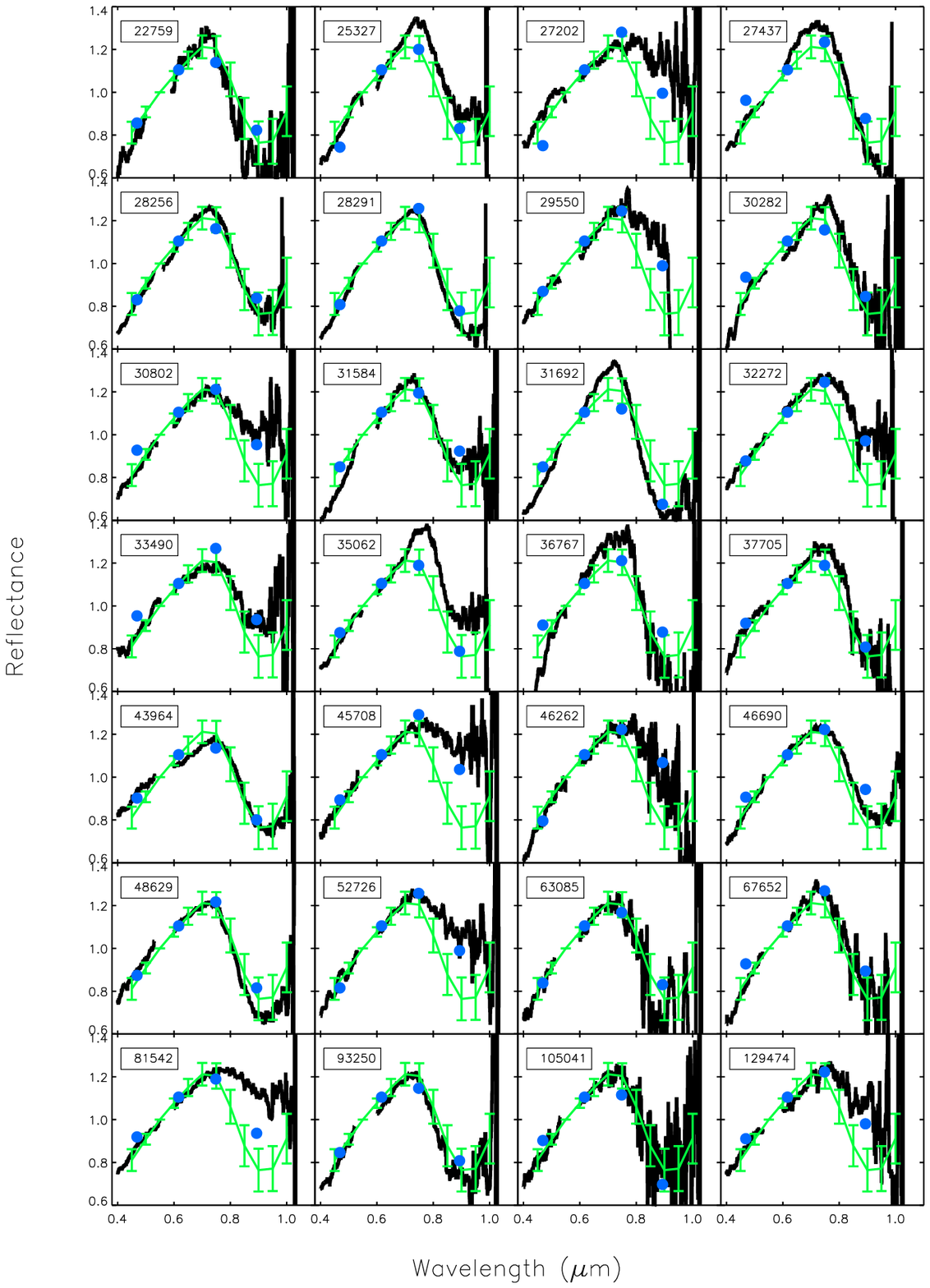} 
Figure \ref{fig_grid} (continued) 
\end{center}
\end{figure}

\clearpage

\end{document}